# Superconducting nanowire single photon detectors fabricated from an amorphous $Mo_{0.75}Ge_{0.25}$ thin film

V. B. Verma[1*], A. E. Lita[1], M. R. Vissers[1], F. Marsili[2], D. P. Pappas[1], R. P. Mirin[1], and S. W. Nam[1]

[1]*National Institute of Standards and Technology, 325 Broadway, Boulder, CO 80305, USA*

[2]*Jet Propulsion Laboratory, 4800 Oak Grove Dr., Pasadena, California 91109, USA*

We present the characteristics of superconducting nanowire single photon detectors (SNSPDs) fabricated from amorphous $Mo_{0.75}Ge_{0.25}$ thin-films. Fabricated devices show a saturation of the internal detection efficiency at temperatures below 1 K, with system dark count rates below 500 counts per second. Operation in a Gifford-McMahon (GM) cryocooler at 2.5 K is possible with system detection efficiencies (SDE) exceeding 20% for SNSPDs which have not been optimized for high detection efficiency.

Superconducting nanowire single photon detectors (SNSPDs) are an important technology for performing experiments in quantum optics requiring high efficiency single photon detectors due to their high timing resolution and low dark count rates.[1-4] Until recently, however, the low quantum efficiency of SNSPDs has limited their use in many applications such as quantum key distribution, quantum state tomography, quantum computing, and loophole-free Bell tests. Since the demonstration of the first SNSPD, NbN and NbTiN have primarily been used as the superconducting materials due to a number of desirable properties such as a high superconducting transition temperature and high critical current density.[5,6] Recently, other materials have been explored for the fabrication of SNSPDs such as TaN,[7] NbSi,[8] amorphous WSi,[9-11] and MoSi.[12] WSi SNSPDs have demonstrated the highest system detection efficiencies (SDE) to date of 93%.[10]

The high efficiencies achievable with WSi are related to the intrinsic photon detection mechanism in an SNSPD.[13-15] In the superconducting state, the supercurrent is carried by Cooper pairs. When the bias current is increased close to the critical current $I_C$ for the nanowire, the absorption of a photon breaks Cooper pairs forming a "hotspot" in which the superconducting gap is suppressed, which in turn leads to the formation of a resistive bridge across the nanowire. In order for the nanowire to become resistive, the width of the nanowire must be comparable to the size of the hotspot formed by the absorbed photon. For a given photon energy, one would expect the size of the hotspot to depend on material parameters such as carrier density and the magnitude of the superconducting gap. The source of the high internal detection efficiency in WSi is thought to be its small superconducting gap energy and lower carrier density compared to NbN, which causes a larger number of quasiparticles to be created per absorbed photon.

Although high SDE is achievable with WSi, a lower carrier density in the material results in significantly lower critical currents compared to NbN-based SNSPDs, which in turn results in a lower signal-to-noise ratio in the measured output pulse and correspondingly larger timing jitter.[5, 6, 9-11] In addition, the superconducting transition temperature ($T_C$) of a WSi SNSPD is typically ~ 3 K. To achieve the highest possible SDE and lowest jitter, operation below 1 K is desirable, which requires the use of specialized and expensive cryogenic systems. The development of low-jitter, high-efficiency SNSPDs operating at 2.5 K, a temperature which is achievable in a closed-cycle Gifford-McMahon (GM) cryocooler, would allow a more widespread application of these detectors. In this work, we present results from SNSPDs fabricated from amorphous $Mo_{0.75}Ge_{0.25}$, which has a higher bulk $T_C$ (6.8 K) than WSi (5 K) but lower than NbN (~ 16 K). We find that the superconducting energy gap is large enough to enable operation of the SNSPDs at 2.5 K, but still low enough to provide a hotspot size adequate to result in saturation of the internal detection efficiency similar to WSi even for relatively wide nanowire geometries between 100 - 150 nm.

The 7.5 nm-thick $Mo_{0.75}Ge_{0.25}$ film was deposited on a Si wafer with 150 nm of thermally grown $SiO_2$. The deposition was performed by DC magnetron cosputtering from separate Mo and Ge targets at room temperature. The amorphous nature of the film was verified by x-ray diffraction. After the deposition, the wafer was transferred to a separate deposition system to be capped with 2 nm of amorphous Silicon to prevent long-term oxidation. The bulk $T_C$ of the MoGe is 6.8 K, which is reduced to 4.4 K for the 7.5 nm-thick film. Fifty nm-thick gold contact pads were deposited and patterned by liftoff prior to the deposition of the MoGe film. After deposition of the thin film, it is etched into 20 μm-wide strips between the contact pads in an $SF_6$ plasma. Nanowire meanders are patterned into the 20 μm strips by electron beam lithography using PMMA and etching in $SF_6$. The area of each nanowire meander is 16 µm × 16 µm. Keyhole-shaped chips are created using a through-wafer etch, which can then be removed from the Si wafer and self-aligned to a single-mode optical fiber. The self-aligned technique allows alignment of the detector to the core of the single-mode fiber to within ± 3 µm.[16] The SNSPDs were mounted inside of an adiabatic demagnetization refrigerator (ADR) for measurement.

Figure 1 shows measurements of the system detection efficiency (SDE) vs. bias current ($I_B$) at temperatures of 250 mK, 1 K, 2 K, and 2.5 K for two SNSPDs at a wavelength of 1550 nm. The nanowire composing the SNSPD in Fig. 1(a) has a width of 110 nm, and a pitch (center-to-center spacing of adjacent nanowires in the meander) of 190 nm. The SNSPD measured in Fig. 1(b) is composed of a 150 nm-wide nanowire with a pitch of 230 nm. The switching currents for the two SNSPDs are 8.5 µA and 11.7 µA, with peak system detection efficiencies of 26% and 30%, respectively for the optimum polarization of the incident light. As shown in Fig. 1(a), the 110 nm-wide nanowire shows saturation of the internal detection efficiency over a bias range between 7 µA and 8.5 µA at 250 mK. The saturation is still evident at a temperature of 1 K. The SNSPD in Fig. 1(b) reaches saturation just before switching to the normal state due to its wider nanowire geometry.[17]

The index of refraction of the 7.5 nm-thick MoGe film with 2 nm Si cap was measured with a spectrophotometer and used to compare numerical simulations of absorption in the patterned film to measured efficiencies. Numerical simulations predict 36% absorption for a nanowire width of 110 nm and pitch of 190 nm, and 37% absorption for a nanowire width of 150 nm and pitch of 230 nm. The predicted absorption for the two SNSPDs differs from measured values of the efficiency by 10% and 7% respectively for the 110 nm and 150 nm-wide nanowires. The magnitude of this discrepancy is similar to that measured with WSi SNSPDs.[9, 10]

The source of the loss requires further investigation, but could potentially be attributed to slight fiber-detector misalignment or internal detection efficiency less than unity despite the saturation in system efficiency.

The dark count rate (DCR) was measured both with the SNSPD coupled to the optical fiber, which we refer to as the system dark count rate (SDCR), as well as without the fiber, which we refer to as the device dark count rate (DDCR). Fig. 2 shows the SDCR vs. bias for both the 110 nm and 150 nm SNSPDs. The SDCR is below 1000 counts per second (cps) for both detectors, and below 500 cps over most of the bias range. Fig. 3 shows the DDCR vs. bias for the same detectors. The DDCR was obtained by averaging for 10 seconds at each point. The DDCR is zero except very close to the switching current. Higher SDCR relative to the DDCR suggests that, like WSi SNSPDs, the MoGe SNSPDs are sensitive to blackbody photons which are coupled into the optical fiber.[10]

Fig. 4 shows a single-trace voltage pulse from the 150 nm-wide SNSPD. The $1/e$ decay time of the pulse is ~ 9 ns. From the decay time of the pulse we obtain a kinetic inductance for the film of ~ 61 pH/square for the 7.5 nm-thick film. The jitter was also measured for this device as a function of temperature with the device biased at 97 % of the switching current and using two room-temperature amplifiers with 56 dB of total gain. A mode-locked 1550 nm laser was used in conjunction with an oscilloscope to construct a histogram of the time difference between the reference pulse from the laser and the SNSPD pulse. We obtained values for the jitter of 60 ps, 72 ps, 102 ps, and 140 ps at temperatures of 250 mK, 1 K, 2 K, and 2.5 K, respectively. Further improvement could potentially be obtained with the use of cryogenic amplifiers.[10]

Finally, it is instructive to compare the performance of MoGe SNSPDs to that of WSi SNSPDs. Both the switching currents and SDE of MoGe SNSPDs are higher than typical WSi SNSPDs at comparable temperatures. In part, the higher switching currents and efficiencies can be attributed to the use of a thicker film (7.5 nm) compared to WSi SNSPDs (4 - 4.5 nm). Regarding the dark count rates, the SDCR for MoGe is less than 500 cps over most of the bias range, while it is generally ~ 1000 cps for WSi SNSPDs. The reduced sensitivity to low-energy blackbody photons may also be attributed to the use of a thicker film and the larger superconducting gap energy. Finally, the MoGe SNSPDs appear to have a faster thermal recovery time compared to WSi detectors, which typically latch to the normal state when the electrical recovery time (determined by the kinetic inductance) is less than 20 ns. Furthermore, the jitter is significantly improved relative to WSi, which can be attributed to the higher switching current and signal-to-noise ratio of the MoGe SNSPDs.

In summary, MoGe shows promise as an alternative amorphous material to WSi for SNSPDs. The amorphous nature of MoGe should allow it to be deposited on a wide range of substrates without degradation of the superconducting material properties such as $T_C$ and critical current density. Efficiencies approaching 100% should theoretically be achievable when embedded inside of an optical stack, similar to WSi SNSPDs.[10] In addition, the jitter is smaller and the recovery times faster with MoGe compared to WSi. At a temperature of 2.5 K, greater than 20% SDE has been achieved with a jitter of 140 ps. Typically the low switching current of WSi (~ 2 µA) at 2.5 K limits the jitter to ~ 500 ps, which limits the usability of the detector in applications requiring high timing resolution. For these reasons MoGe SNSPDs are expected to be useful detectors for many applications when operation in a closed-cycle GM cryocooler at 2.5 K is required.

**Figures**

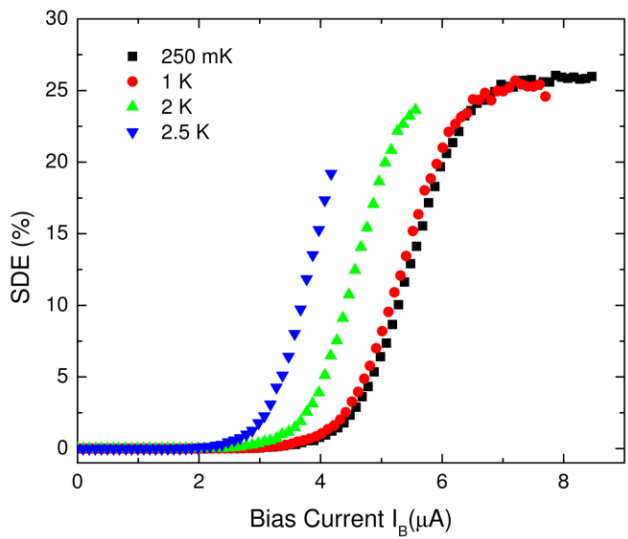

(a)

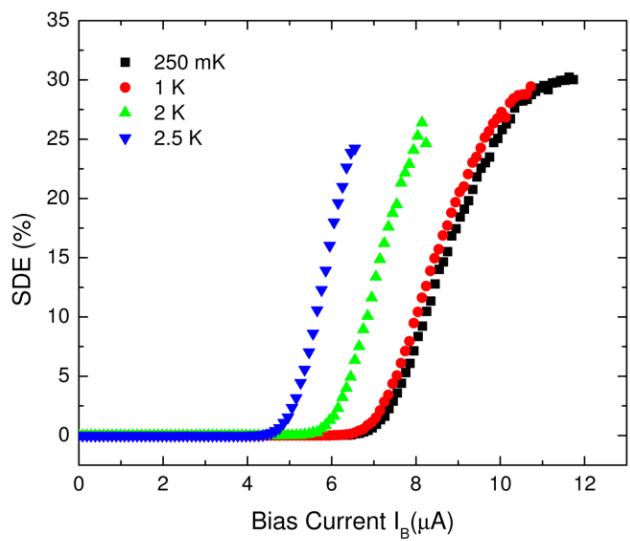

(b)

Fig. 1 System detection efficiency (SDE) vs. bias current ($I_B$) for SNSPDs having nanowire widths of (a) 110 nm and (b) 150 nm. Measurements were performed at temperatures of 250 mK, 1 K, 2 K, and 2.5 K.

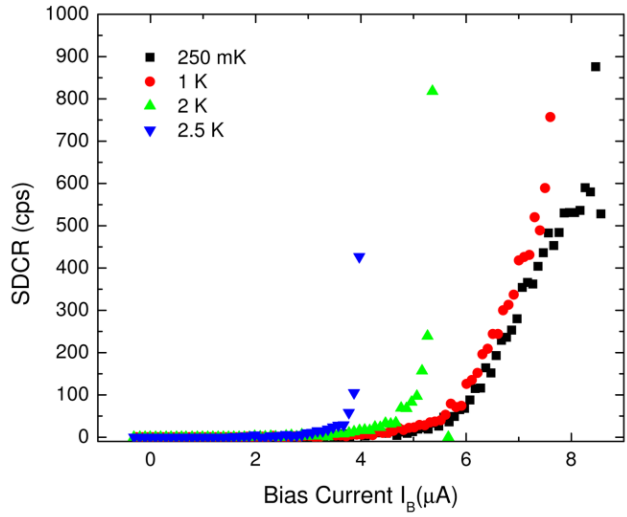

(a)

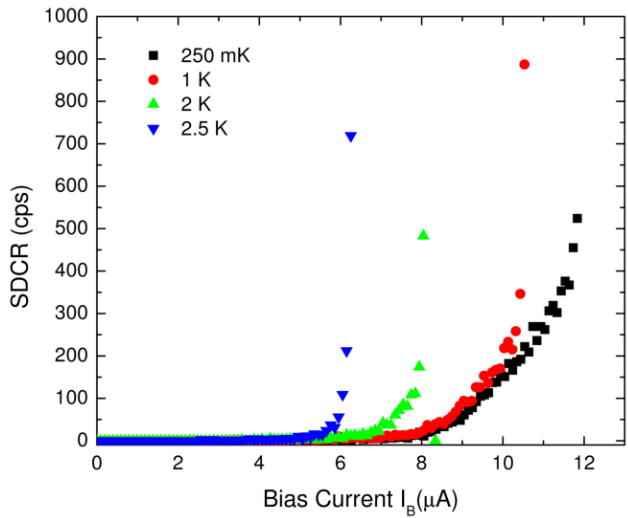

(b)

Fig. 2 System dark count rate (SDCR) vs. bias current $I_B$ for SNSPDs having nanowire widths of (a) 110 nm and (b) 150 nm. Measurements were performed at temperatures of 250 mK, 1 K, 2 K, and 2.5 K.

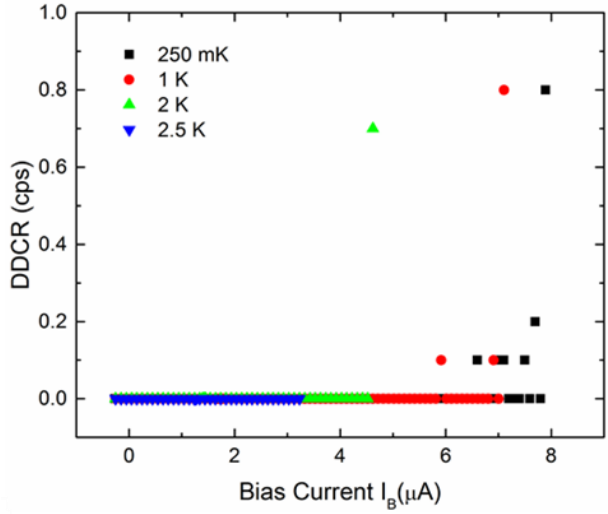

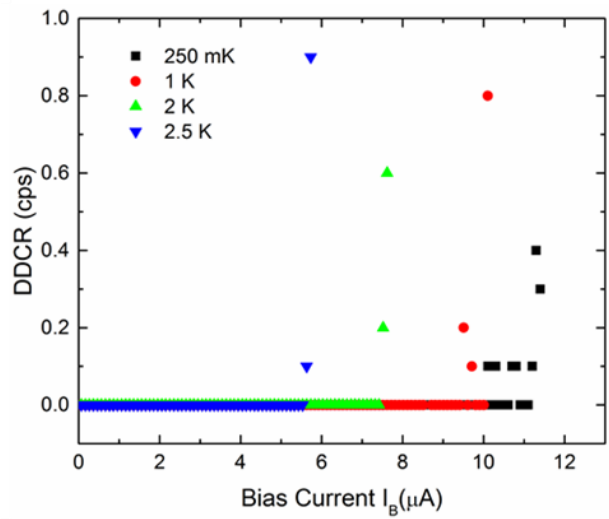

Fig. 3 Device dark count rate (DDCR) vs. bias current $I_B$ for SNSPDs having nanowire widths of (a) 110 nm and (b) 150 nm. Measurements were performed at temperatures of 250 mK, 1 K, 2 K, and 2.5 K.

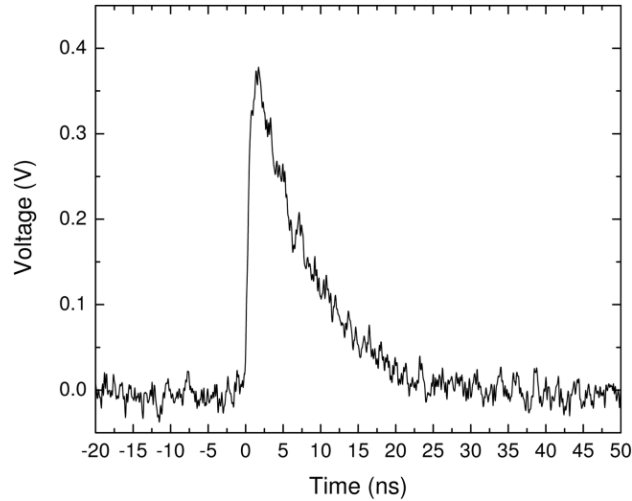

Fig. 4 Voltage pulse from the 150 nm-wide SNSPD.

**References**

1   H. Takesue, S. W. Nam, Q. Zhang, R. H. Hadfield, T. Honjo, K. Tamaki, and Y. Yamamoto, Nat. Photonics **1**, 343 (2007).

2   J. Chen, J. B. Altepeter, M. Medic, K. F. Lee, B. Gokden, R. H. Hadfield, S. W. Nam, and P. Kumar, Phys. Rev. Lett. **100**, 133603 (2008).

3   C. Clausen, I. Usmani, F. Bussieres, N. Sangouard, M. Afzelius, H. de Ridematten, and N. Gisin, Nature London **469**, 508 (2011).

4   M. J. Stevens, B. Baek, E. A. Dauler, A. J. Kerman, R. J. Molnar, S. A. Hamilton, K. K. Berggren, R. P. Mirin, and S. W. Nam, Opt. Express **18**, 1430 (2010).

5   G. N. Gol'tsman, O. Okunev, G. Chulkova, A. Lipatov, A. Semenov, K. Smirnov, B. Voronov, and A. Dzardanov, Appl. Phys. Lett. **79**, 705 (2001).

6   S. Miki, T. Yamashita, H. Terai, and Z. Wang, Optics Express **21**, 10208 (2013).

7   A. Engel, A. Aeschbacher, K. Inderbitzin, A. Schilling, K. Il'in, M. Hofherr, M. Siegel, A. Semenov, and H. -W. Hübers, Appl. Phys. Lett. **100**, 062601 (2012).

8   S. N. Dorenbos, P. Forn-Dìaz, T. Fuse, A. H. Verbruggen, T. Zijlstra, T. M. Klapwijk, and V. Zwiller, Appl. Phys. Lett. **98**, 251102 (2011).

9   B. Baek, A. E. Lita, V. Verma, and S. W. Nam, Appl. Phys. Lett. **98**, 251105 (2011).

10  F. Marsili, V. B. Verma, J. A. Stern, S. Harrington, A. E. Lita, T. Gerrits, I. Vayshenker, B. Baek, M. D. Shaw, R. P. Mirin, and S. W. Nam, Nat. Photonics **7**, 210 (2013).

[11] V. B. Verma, F. Marsili, S. Harrington, A. E. Lita, R. P. Mirin, and S. W. Nam, Appl. Phys. Lett. **101**, 251114 (2012).

[12] Y. P. Korneeva, M. Y. Mikhailov, Y. P. Pershin, N. N. Manova, A. V. Divochiy, Y. B. Vakhtomin, A. A. Korneev, K. V. Smirnov, A. Y. Devizenko, and G. N. Goltsman, arXiv:1309.7074 [cond-mat.supr-con]

[13] L. N. Bulaevskii, M. J. Graf, and V. G. Kogan, Phys. Rev. B **85**, 014505 (2012).

[14] A. D. Semenov, G. N. Gol'tsman, and A. A. Korneev, Physica C **351**, 349 (2001).

[15] A. Semenov, A. Engel, H. -W. Hübers, K. Il'in, and M. Siegel, Eur. Phys. J. B **47**, 495(2005).

[16] A. J. Miller, A. E. Lita, B. Calkins, I. Vayshenker, S. M. Gruber, and S. W. Nam, Opt. Express **19**, 9102 (2011).

[17] F. Marsili, F. Bellei, F. Najafi, A. E. Dane, E. A. Dauler, R. J. Molnar, and K. K. Berggren, Nano Lett. **12**, 4799 (2012).